\documentclass[twocolumn]{aastex631}

\usepackage{amsmath}

\begin{document}

\title{Magnetized winds of M-type stars and star-planet magnetic interactions: uncertainties and modeling strategy}

\author[0000-0002-2916-3837]{Victor Réville}
\affiliation{IRAP, Université Toulouse III - Paul Sabatier, CNRS, CNES, Toulouse, France}

\author[0000-0001-9969-2884]{Jamie M. Jasinski}
\affiliation{NASA Jet Propulsion Laboratory, California Institute of Technology, Pasadena, CA, USA}

\author[0000-0002-2381-3106]{Marco Velli}
\affiliation{NASA Jet Propulsion Laboratory, California Institute of Technology, Pasadena, CA, USA}

\author[0000-0002-9630-6463]{Antoine Strugarek}
\affiliation{D\'epartement d'Astrophysique/AIM,
CEA/IRFU, CNRS/INSU, Univ. Paris-Saclay \& Univ. de Paris, 
91191 Gif-sur-Yvette, France}

\author[0000-0002-1729-8267]{Allan Sacha Brun}
\affiliation{D\'epartement d'Astrophysique/AIM,
CEA/IRFU, CNRS/INSU, Univ. Paris-Saclay \& Univ. de Paris, 
91191 Gif-sur-Yvette, France}

\author[0000-0001-6102-7563]{Neil Murphy}
\affiliation{NASA Jet Propulsion Laboratory, California Institute of Technology, Pasadena, CA, USA}

\author[0000-0002-7628-1510]{Leonardo H. Regoli}
\affiliation{Johns Hopkins Applied Physics Laboratroy, Laurel, MD, USA}

\author[0000-0003-4039-5767]{Alexis Rouillard}
\affiliation{IRAP, Université Toulouse III - Paul Sabatier, CNRS, CNES, Toulouse, France}

\author[0000-0002-6114-0539]{Jacobo Varela}
\affiliation{Institute for Fusion Studies, Department of Physics, University of Texas at Austin, Austin, Texas 78712, USA}

\begin{abstract}
M-type stars are the most common stars in the universe. They are ideal hosts for the search of exoplanets in the habitable zone (HZ), as their small size and low temperature make the HZ much closer in than their solar twins. Harboring very deep convective layers, they also usually exhibit very intense magnetic fields. Understanding their environment, in particular their coronal and wind properties, is thus very important, as they might be very different from what is observed in the solar system. The mass loss rate of M-type stars is poorly known observationally, and recent attempts to estimate it for some of them (TRAPPIST-1, Proxima Cen) can vary by an order of magnitude. In this work, we revisit the stellar wind properties of M-dwarfs in the light of the latest estimates of $\dot{M}$ through Lyman-$\alpha$ absorption at the astropause and slingshot prominences. We outline a modeling strategy to estimate the mass loss rate, radiative loss and wind speed, with uncertainties, based on an Alfvén wave driven stellar wind model. We find that it is very likely that several TRAPPIST-1 planets lie within the Alfvén surface, which imply that these planets experience star-planet magnetic interactions (SPMI). We also find that SPMI between Proxima Cen b and its host star could be the reason of recently observed radio emissions. 
\end{abstract}



\section{Introduction} \label{sec:intro}

The characterization of cool M-dwarf astrospheres is of primary importance to the study of the environment of their orbiting planetary systems. M-dwarf stars are very active, creating large average magnetic fields in their deep convective envelope, up to the kilo gauss scale \citep{ReinersBasri2010, Moutou2017, Reiners2022}. Even slow rotators have been recently shown to be able to create a few hundred gauss fields, largely above the expected strength from their Rossby number \citep{Lehmann2024}. The correlation between the field strength and coronal heating of solar-like stars is moreover well established \citep[see, e.g][]{Gudel2004}, which directly translates into more wind power for higher stellar magnetic fields. The indirect observations of stellar winds using the Lyman-$\alpha$ (Ly$\alpha$) absorption signatures of the astrospheres \citep{Wood2005} provides a power law relating the X-ray power of the corona and the mass loss rate of cool stars. This relation has been reassessed recently using new measurement of M-dwarf and slingshot prominences detected in very active stars \citep{Wood2021, Jardine2019}. 

The mass loss rate is, however, the only true constraint that we possess on the properties of distant stellar winds. Many parameters that are relevant to the interaction of exoplanets with the stellar environment, are not observable remotely. For example, knowing the mass loss rate and the stellar wind speed provides the dynamic pressure, but the wind speed is unknown for all main sequence stars but the Sun. The dynamic pressure is essential to compute the magnetospheric stand-off distance, in addition to knowing the planetary magnetic field \citep[see, e.g.][]{Varela2022a, Varela2022b}. Moreover, Extreme Ultraviolet (EUV) emissions are difficult to constrain, being mostly absorbed in the interstellar medium. Yet, the EUV flux is among the main factors impacting the atmospheric escape rate of exoplanets \citep[][]{Lecavelier2007, Bourrier2017b}.

In the particular context of M-dwarfs, for which potentially habitable planets lie very close to the star, another interesting question arises. Given their very intense magnetic field, one can wonder whether planets in the habitable zone (HZ) also orbit within the Alfvén surface of the star, i.e., at a distance where the planet can have a magnetic influence on its host. Star-planet magnetic interactions \citep[SPMI, see][]{Strugarek2015,Strugarek2019}, can influence the secular orbital evolution of systems \citep{Strugarek2014}, and are expected to generate intense energy exchange between the planet and the star \citep{Strugarek2022}.

Recently, strong radio emission coming from Proxima Centauri (hereafter Proxima Cen) have been detected \citep{PerezTorres2021}. With a semi-major axis of $\sim 74 R_\odot$ \citep{Anglada2016}, Proxima Cen b has been proposed as a possible source of these radio emissions, through a cyclotron maser emission process analogous to the Io-Jupiter interaction \citep{Zarka1996}. Previous simulations of the Proxima Cen system have claimed that planet b should lie outside the Alfvén surface \citep{Garraffo2016, Kavanagh2021}. While both based on the Alfvén wave driven solar wind model AwSoM \citep{vanderHolst2014}, they use various magnetic field constraints on the inner boundary and obtain very different mass loss rates, very well above the non-detection limit of \citet{Wood2001} in the case of \citet{Garraffo2016}. The same kind of discrepancies are found within studies of the TRAPPIST-1 system, in which seven close-in planets have been discovered \citep{Gillon2016, Gillon2017}. For this particular system, the predictions made with the AwSoM code have either most of the orbiting planets within the Alfvén surface \citep{Garraffo2017}, or none \citep{Dong2018}. The resulting mass loss rates are also very different, varying between $1 \dot{M}_\odot$ \citep{Garraffo2017} and $0.1 \dot{M}_\odot$ \citep{Dong2018}, which likely explains part of these differences. 

In this work, we propose to reevaluate the stellar wind modeling strategy guided by the updated measurements of \citet{Wood2021}. We perform 9 simulations using the Alfvén wave driven solar wind model WindPredict-AW, varying the main input parameters. In section \ref{sec:wood}, we discuss briefly the observed mass loss rate constraints. Section \ref{sec:num} describes the fundamentals of the WindPredict-AW model, and the modifications brought to the code for the case of M-dwarf. Then, section \ref{sec:trap} shows the results of the simulations and the various channels into which the input energy is distributed: radiation, mass loss, and wind acceleration. We show that the characteristics of the turbulence plays a key role in the terminal wind speed. We highlight the stellar wind conditions that should be characteristic of the TRAPPIST-1 system, and find in Section \ref{sec:spi}, that inner TRAPPIST-1 planets should orbit within the Alfvén surface, giving birth to SPMI. We also reassess the situation of Proxima Cen b, and find that it could well be the source of strong radio emissions. Finally, section \ref{sec:xuv}, we explore ways to assess the X-ray and UV spectra of M-dwarfs, and use them to constrain our simulations.

\section{Mass loss rate observations}
\label{sec:wood}

\begin{figure}
    \centering
    \includegraphics[width=3.5in]{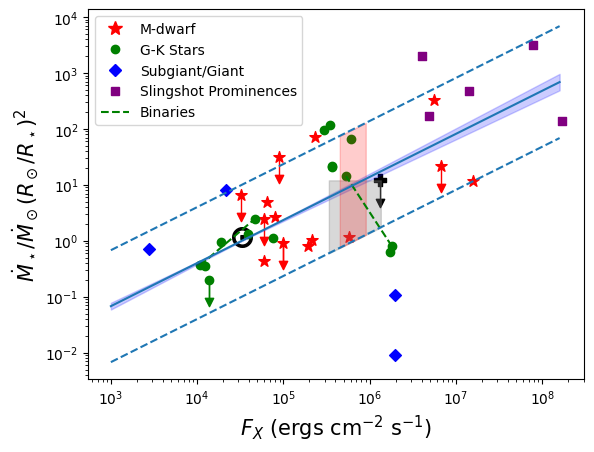}
    \caption{\citet{Wood2021} $\dot{M}-F_X$ law with updated measurements from M-dwarfs and slingshot prominences. The blue line represents the fitted law (\ref{eq:wood_law}). The dashed lines, 1/10 and 10 times the fitted law, encompass most of the 36 data points, with only a few outliers on the subgiant / giant branch. The filled region in red and grey correspond to the mass loss rate estimates for TRAPPIST-1 and Proxima Cen respectively. The black cross symbol is the upper bound for Proxima Cen (all down arrows represent upper bounds).}
    \label{fig:wood_law}
\end{figure}

The \citet{Wood2021} $F_X-\dot{M}$ relationship can be expressed as follows, 

\begin{equation}
\frac{\dot{M}_\star}{R_\star^2} =  \frac{\dot{M}_{\odot}}{R_\odot^2} \left(\frac{F_{X, \star}}{F_{X, \odot}} \right)^{0.77 \pm 0.04},
\label{eq:wood_law}
\end{equation}
where $F_X$ is the X-ray flux and $\dot{M}$ the mass loss rate. Figure \ref{fig:wood_law} shows the 35 measured mass loss {rates} through the astrospheric detection method \citep{Wood2005} and the slingshot prominence technique \citep{Jardine2019}. The Sun is added with a $\odot$ symbol, while M-dwarf and subgiant/giant stars are identified with red stars and blue diamonds respectively. Main sequence G-K stars are represented with green dots. In this plot, we chose a solar X-ray flux $F_{X, \odot} = 10^{27.3}$ erg/s, and a mass loss rate of $\dot{M}_{\odot}=1.5 \times 10^{12}$ g/s, which is slightly higher than the value used in \citet{Wood2021}, and consistent with the activity maximum \citep{Judge2003}.

The blue line shows equation (\ref{eq:wood_law}), the uncertainties in the exponent being represented in blue shades. One can see, however, that the spread remains much larger than these uncertainties, and we added two curves multiplying by a factor 0.1 and 10 the fitted law, shown in dashed blue color. These two curves do cover most of the data points, except two points in the lowest part of the diagram, corresponding to the giants stars $\lambda$ And and DK UMa \citep[see][for more details]{Wood2021}.

In the following, we will aim to model the wind of two specific M-dwarf, TRAPPIST-1 and Proxima Cen, with mass loss {rates} compatible with this observationally constrained scaling law. Given the uncertainties illustrated in Figure \ref{fig:wood_law}, we shall consider that a mass loss rate of one order of magnitude lower or higher than the one predicted by Wood law's is perfectly plausible. The range of expected mass loss {rates} is highlighted in grey shades for Proxima Cen given the quiet X-ray luminosity $L_X = 4-16 \times 10^{26}$ erg/s \citep{Haisch1990}. It is bounded by the non-detection limit found by \citet{Wood2001} and shown with the black cross. The equivalent mass loss rate range is shown in red shades for TRAPPIST-1, for measured $L_X = 4-8 \times 10^{26}$ erg/s \citep{Wheatley2017}.

\section{Numerical Model}
\label{sec:num}

We use the WindPredict-AW model \citep{Reville2020ApJS, Parenti2022}, based on the PLUTO code \citep{Mignone2007}. The implementation is very similar to the one described in \citet{Reville2021, Reville2022}, where we use a chromospheric boundary condition with a transition region. The boundary conditions are extensively discussed in \citet{Parenti2022}, the only difference being that the temperature is held fixed at the chromospheric inner boundary. 

The model solves the ideal MHD equations, assuming that the wind driving is solely due to the turbulent heating and wave pressure of Alfvén wave packets propagating from the surface of the Sun. We neglect any heating contribution from impulsive events such as nanoflares \citep{Parker1988}, that would require resistivity. The ideal MHD system is thus complemented by two equations for Alfvén wave transport and dissipation. These equations read: 

\begin{equation}
    \frac{\partial \mathcal{E}^\pm}{\partial t} + \nabla\cdot\left([\boldsymbol{v}\pm\boldsymbol{v_A}]\mathcal{E}^\pm\right) = -\frac{\mathcal{E}^\pm}{2}\nabla\cdot\boldsymbol{v}- (\mathcal{R}^\pm \mathcal{E}^\pm- \mathcal{R}^{\mp} \mathcal{E}^\mp)-Q_w^\pm,
\end{equation}
where $\mathcal{E}^\pm = \rho |z^\pm|^2/4$ is the energy density of Alfvén wave packets defined by the Elsässer variables 

\begin{equation}
    z^\pm = \delta v \mp \mathrm{sign}(B_r) \frac{\delta b}{\sqrt{4 \pi \rho}}.
\end{equation}

The turbulence dissipation term is defined as:

\begin{equation}
    Q_w^{\pm} = \frac{\rho}{8}\frac{|z^{\pm|^2}}{\lambda}|z^\mp|,
\end{equation}

where $\lambda = \lambda_\star \sqrt{\langle B_\star \rangle/|B|}$, is the turbulence correlation length scale and $\mathcal{R^\pm}$ is the reflection coefficient of Alfvén waves in the expanding solar wind and is defined

\begin{equation}
    \mathcal{R}^\pm = |V_{A,r} \nabla_r \ln \sqrt{\rho}| \times \mbox{MAX} \left(0, 1-\frac{\mathcal{E}^\mp}{\mathcal{E}^\pm}\right). 
    \label{eq:reflection}
\end{equation}

This represents the main modification compared to previous versions of WindPredict-AW. The form of the reflection term is inspired by the incompressible equations of Alfvén waves \citep[see, e.g.,][]{Velli1993}, and is very close to similar models \citep{vanderHolst2014, Shi2023}. This term makes the reflection proportional to the density (or Alfvén speed) gradients in the solar wind. Around the transition region, the inverse reflection timescale $|V_{A,r} \nabla_r \ln \sqrt{\rho}|$ is however very large due to the magnetic field amplitude, and thus it is limited by the ratio of the two Elsässer population, up to a roughly balanced turbulence ($|z^+| \sim |z^-|$). The max term also guarantees that only the dominant wave population undergo reflection, which ensures a global conservation of energy in the system: the reflection loss and gain term compensates exactly, while the dissipation term is converted into heat by the energy conservation equation.

The energy sources and sinks are $Q = Q_w - Q_r - Q_c$, the sum of wave energy dissipation $Q_w = Q_w^+ +Q_w^-$, the radiative losses $Q_r= n^2 \Lambda(T)$, where $\Lambda$ is computed as a fit to the CHIANTI database radiative loss function for solar coronal abundances \citep[version 10.0.1,][]{Dere1997, DelZanna2021, Schmelz2012}.  

A few modifications have been brought to the thermal conduction sink term $Q_c$, in order to account for much larger closed loops, compared to the solar case. It is implemented as:
\begin{equation}
    Q_c = \nabla\cdot(\alpha\boldsymbol{q}_s + (1-\alpha)\boldsymbol{q}_p),
\end{equation}
where $\alpha=1/(1+(r-R_\odot)^2/(r_{coll}-R_\odot)^2)$. We changed here the radial dependence of the parameter $\alpha$ from a power law decrease of index $-4$ to $-2$. The height at which the transition between the collisional and collisionless regimes remains $r_{coll}=5R_\odot$, but the parabolic Braginskii thermal diffusion, $\boldsymbol{q}_s=-\kappa_0T^{5/2} (\hat{\mathbf{b}} \cdot \nabla T) \hat{\mathbf{b}}$ is still significant up to $20 R_{\odot}$. We then set the collisionless electron heat flux $\boldsymbol{q}_p = 4 p_{th} (\hat{\mathbf{b}} \cdot \boldsymbol{v}) \hat{\mathbf{b}}$, for which the prefactor has been raised from $3/2$ to $4$ compared to previous papers, in agreement with the analytical calculations of \citet{Hollweg1976}. 

\begin{figure*}
    \centering
    \includegraphics[width=6.5in]{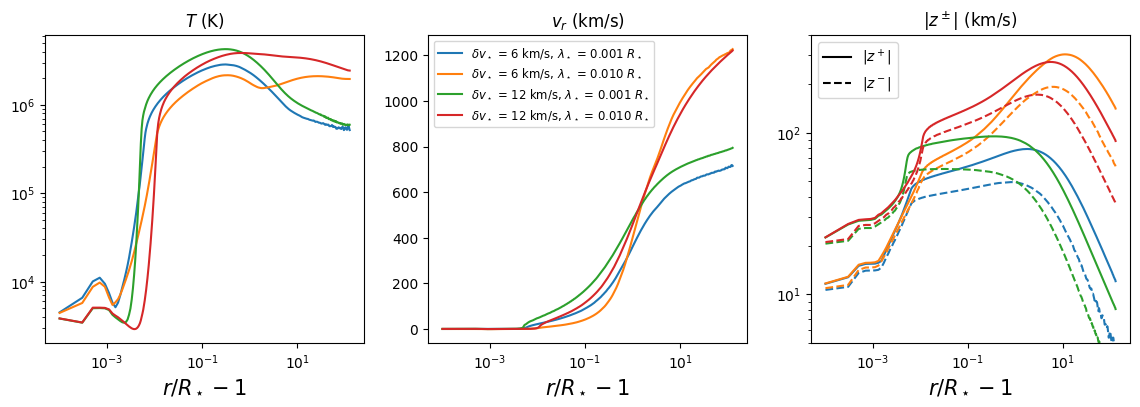}
    \caption{Profiles of the coronal temperature, radial wind speed and wave populations along an open field regions for four simulations. }
    \label{fig:profiles}
\end{figure*}

The base density and temperature are set to $n_\star = 2\times 10^{13}$ cm$^{-3}$ and $T_\star=3000$ K. The base density is about 10 times higher than for similar typical solar wind simulation, while the temperature is the characteristic photospheric temperature for M-dwarf. The stellar atmosphere hence goes through a transition region where the temperature jumps up to a few $10^6$ K. We use a 2.5D spherical grid with $[896, 256]$ points to cover the domain $r \in [1,130 R_\star]$ and $\theta \in [0, \pi]$. The grid is strongly refined at the inner boundary, as well as near the temperature maximum between 1 and $4R_\star$. Moreover, some numerical adjustments are necessary to properly describe the energy transfer in the transition region. For this, we use the technique of \citet{Lionello2009}, which sets a cutoff temperature under which the product of the thermal conduction and the radiative losses are kept constant. We set the cutoff temperature to $T_c = 4 \times 10^5$K.  This comes down to multiplying the terms $Q_r$ and $Q_w$ by a factor $T^{5/2}/T_c^{5/2}$, which ensures that the integrated energy terms are the same as with a fully resolved transition region \citep[see][for more details]{Johnston2020}. This will be very important for the following analysis, as shown in section \ref{sec:trap} and section \ref{sec:xuv}.

The remaining input parameter are thus the initial and inner boundary magnetic field, the transverse velocity perturbation $\delta v_{\star}$, and the turbulence correlation length $\lambda_{\star}$ at the base of the domain. In the next section, we will vary the two latter parameters to model the wind of one specific M-dwarf system: TRAPPIST-1.

\begin{table}
\caption{\label{tab:char} Observational parameters of TRAPPIST-1}
\begin{ruledtabular}
\begin{tabular}{|l|c|c|}
        Qty & Value & Ref\\
        \hline
        $R_{\star}$ & $0.121 R_{\odot}$ & \citet{VanGrootel2018}\\
        $M_{\star}$ & $0.089 M_{\odot}$ & \citet{VanGrootel2018}\\
        $Bf$ & $600$ G & \citet{ReinersBasri2010}\\
        $P_{\mathrm{rot}}$ & $3.3$ d & \citet{Luger2017}\\
        $F_X$ & $4-8 \times 10^{26}$ erg/s & \citet{Wheatley2017}        
        \end{tabular}
\end{ruledtabular}
\end{table}

\section{Application to the TRAPPIST-1 system}
\label{sec:trap}

\subsection{Simulation Results}
Observational constraints of the TRAPPIST-1 system are summarized in Table \ref{tab:char}. The radius and mass of the host star set the rest of the normalization triplet ($n_\star, R_\star, V_\star)$, the normalization length $R_\star$ and velocity $v_{\mathrm{kep}, \star} = \sqrt{GM_\star/R_\star}$. \citet{ReinersBasri2010} have measured the magnetic flux of TRAPPIST-1 through Zeeman broadening $Bf = 600$G, where f is the magnetic filling factor. Without any Zeeman Doppler inversion (Donati \& Moutou, private communication), we are not able to recover the energy partition of the magnetic field decomposed on spherical harmonics, and we thus assume a simple dipole with $\langle B_r \rangle =  600$G. This is reasonable as observed M-dwarfs magnetic fields do show a strong dipolar component \citep[see, e.g.,][]{Morin2008,Lehmann2024}. Moreover, even though more complex field can lead to smaller Alfvén surfaces \citep{Reville2015}, the dipole component is what sets the properties of the wind to zeroth order \citep{Finley2017, Finley2018}. These observed dipoles are however not generally aligned with the stellar rotation axis, which leads to time dependent structure of the magnetic sectors. These aspects are out of the scope of the present paper. Moreover, given the weak surface rotation velocity of TRAPPIST-1 and Proxima Cen, we do not expect rotation to have a strong influence on the final solution \citep[such as magneto-centrifugal effects, see][]{Reville2015}, and we thus consider that our 2.5D simulation can represent any inclination angle with respect to the ecliptic plane. The rotation period is accounted for in the rotating frame, with a rotation velocity at the stellar surface $v_{rot} / v_{esc} = 3.5 \times 10^{-3}$ for TRAPPIST-1. 

With the boundary magnetic field set, we now vary the input wave forcing $\delta v_\star$ and the turbulence correlation length scale $\lambda_\star$. We perform 8 simulations with increasing value of $\delta v_\star$ and two values for the turbulence correlation scale. Table \ref{tab:res} summarizes the parameters and results of the 8 simulations. Note that for $\delta v_\star=3$ km/s and $\lambda_\star$=0.01$R_\star$, the code was not able to maintain high temperature regimes in the coronal domain. Tests suggest that further numerical developments are required to reach a physically meaningful steady-state for this set of parameters.

\begin{table}
\caption{\label{tab:res} Simulation parameters and results}
\begin{ruledtabular}
\begin{tabular}{|r|r|r|r|r|}
        $\delta v_\star$ (km/s) &  $\lambda_\star ({R_\star})$ & $L_p$ (erg/s) & $\dot{M}/\dot{M}_{\odot}$ & $\langle R_{A,f} \rangle/R_\star$\\
        \hline
        3  &  0.001 & 1.28e28 & 0.085 & 85\\ 
        6  &  0.001 & 6.36e28 & 0.64  & 52 \\    
        12 &  0.001 & 2.82e29 & 4.4   & 36 \\    
        24 &  0.001 & 1.24e30 & 31.9   & 22\\    
        3  &  0.01  & -       & -     & -  \\    
        6  &  0.01  & 5.36e28 & 0.25  & 66 \\    
        12 &  0.01  & 2.41e29 & 1.79  & 45 \\    
        24 &  0.01  & 1.13e30 & 13.6  & 29 \\  
        \end{tabular}
\end{ruledtabular}
\end{table}

\begin{figure*}
    \centering
    \includegraphics[width=5.5in]{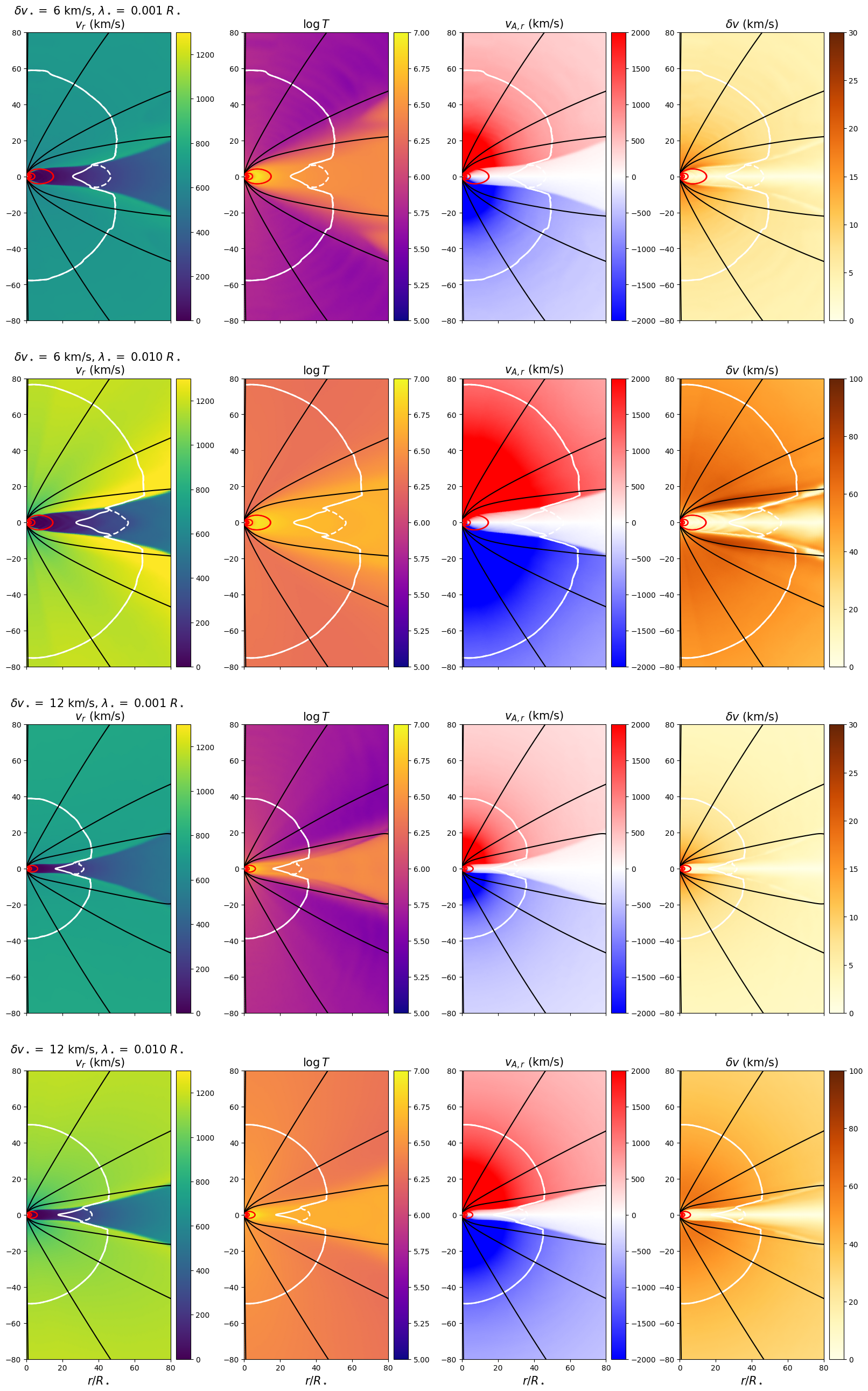}
    \caption{Summary of four cases with the radial velocity, logarithm of temperature, radial Alfvén speed ($B_r/\sqrt{4 \pi \rho}$) and velocity perturbation $\delta v$. The (fast) Alfvén surface is shown in (dashed) white on top of each panel, while open (closed) magnetic field lines are shown in black (red). The $\delta v$ color scale varies for cases with different $\lambda_\star$.}
    \label{fig:viz}
\end{figure*}

Figure \ref{fig:profiles} shows profiles of the temperature, radial wind speed and Alfvén wave amplitudes for a subset of four simulations, with $\delta v_\star=6, 12$ km/s and $\lambda_\star=0.01, 0.001 R_\star$. The temperature of the stellar wind goes from $3000$ K up to $1-4$ MK depending on the simulation parameters. The maximum temperature is logically a growing function of the input $\delta v_\star$. One can notice also that high temperatures are sustained much further away from the star for $\lambda_\star = 0.01 R_\star$. This can be understood looking at the profiles of the Elsässer amplitudes. While starting with similar values, both wave species are dissipated lower in the corona for the highest value of $\lambda_\star$. As a result, their amplitude remain much below $100$ km/s, while they can reach $300$ km/s for $\lambda_\star = 0.01 R_\star$. This generally also explains the velocity profiles, which are strongly dependent on the value of $\lambda_\star$. Fastest winds are obtained for low values of $\lambda_\star$, because of the ponderomotive force (or wave pressure), that acts as an external momentum source for the wind speed. The stellar “fast” wind components thus range between $\sim 700$ km/s and $\sim 1200$ km/s, which is much higher than the fast solar wind. 

Finally, the left panel of Figure \ref{fig:profiles}, shows that in the chromosphere, the turbulence is almost balanced, i.e., $|z^+| \sim |z^-|$, due to the strong reflection below the transition region (see equation \ref{eq:reflection}). Consequently, part of the wave energy injected at the boundary in the outgoing wave re-enters the domain with the oppositely propagating wave population. We use an outflow boundary condition for the inward propagating wave at the inner boundary, and the effective energy injected is thus a result of the simulation. In Table \ref{tab:res}, we report the effective input Poynting luminosity (see section \ref{subsec:energetics}) resulting from this boundary condition.

The four panels of Figure \ref{fig:viz} shows 2D views of each simulation, the radial wind speed, the logarithm of the temperature, the radial Alfvén speed and the velocity perturbations. The global structure of each simulation is very reminiscent of typical solar wind simulations, with open solar wind structures at the two poles of the dipole and closed loops followed by a current sheet and plasma sheet around the equator. The wind speed is modulated by the magnetic structure, with higher wind speed coming from coronal holes and lower wind speed around the current sheet.

Closed coronal loops, or helmet streamers, are nonetheless much larger than any solar structure, especially for cases with low $\delta v_\star$. In the case with $\delta v_\star = 6$ km/s, the tip of these streamers goes beyond $20 R_\star$, which necessitated some changes in our numerical techniques (see section \ref{sec:num}). From the first column, we can observe that the fastest wind speed is between $600$ and $800$ km/s for the three simulations with $\lambda_\star = 0.001 R_\star$, while it reaches $1200$ km/s for $\lambda_\star = 0.01 R_\star$.

\begin{figure}
    \centering
    \includegraphics[width=3.5in]{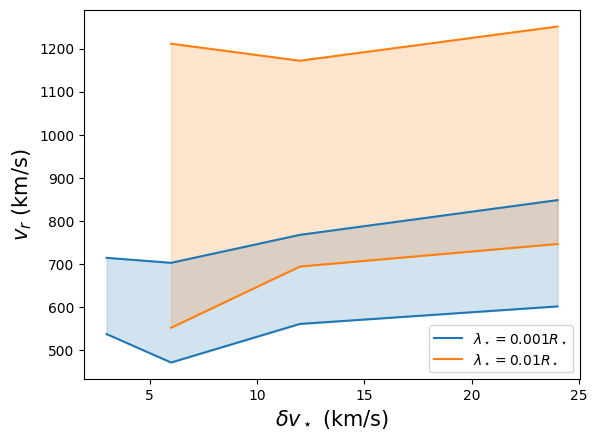}
    \caption{Fast and slow wind speed as a function of $\delta v_\star$ and values of $\lambda_\star$. The fast wind speed increase strongly with increasing value of $\lambda_\star$, reaching some $1200$ km/s for $\lambda_\star=0.01$, with a much weaker dependance on $\delta v_\star$.}
    \label{fig:speed_range}
\end{figure}

We plot in Figure \ref{fig:speed_range}, the characteristic slow and fast wind speed and the subsequent range of possible wind speed for each set of simulations. We computed the area weighted average speed of each wind component at a distance of $100 R_\star$, where we assume that the terminal velocity is reached. This plot shows even more clearly the strong dependency of the wind speed to $\lambda_\star$. The wind speed is a growing function of $\lambda_\star$ and, in most cases, also a growing function $\delta v_\star$ (see Figure \ref{fig:speed_range}). Nevertheless, we observe a non-monotonic behavior of the wind speed at low $\delta v_\star$. This is a reminder that the dynamics of Alfvén wave turbulent transport is non-trivial. In essence, because of the lower heating for $\delta v_\star = 3$ km/s, the wind is very tenuous, which in return yields higher Alfvén speeds and stronger amplification of Alfvén waves in the corona, able to accelerate the wind to higher velocities. 

In the solar corona, close to the surface, closed regions and sources of slow wind are hotter than open field areas, due to higher expansion rates that increase heating. This trend reverse further away where the fast wind becomes hotter than the slow wind \citep[see][for simulations of the solar cycle]{Hazra2021}. For the TRAPPIST-1 system, we also observe hotter slow wind close-in, but the slow wind remains hotter than the fast wind throughout the simulation domain. It is nonetheless likely that this transition is only displaced further away due to the very intense magnetic field of M-dwarf. There are also more contrast in temperature between the slow and fast wind depending on the value of $\lambda_\star$. We recover the property seen in Figure \ref{fig:profiles}, that MK temperature are sustained to further distances with higher $\lambda_\star$. This can be understood rather straightforwardly. The turbulence correlation length scale parameter controls the scale height of turbulent heating in the stellar wind, and lower values will tend to concentrate the heating in the chromosphere and very low corona. This will not only load the stellar wind to higher densities but also dissipate Alfvén waves much more efficiently close to the star. As such, as shown in the last left most panels of Figures \ref{fig:profiles} and \ref{fig:viz}, there is much less wave energy to heat the solar corona at a further distance from the star.
 
\subsection{Energetics}
\label{subsec:energetics}

Let us now discuss the global energetics of the simulations. Assuming steady state, the energy equation can be written as: 

\begin{equation}
    \begin{aligned}
    \nabla \cdot \mathbf{F}_E & =\\
    \nabla \cdot \left[\rho \mathbf{v} \left(\frac{1}{2}v^2 + \frac{c_s^2}{\gamma-1} - \frac{GM_{\star}}{r} \right) + \mathbf{F}_p + \mathbf{F}_c \right] &= -Q_r. \\
    \end{aligned}  
\end{equation}

Where $\mathbf{F}_E$ is the total energy flux, $\mathbf{F}_p$ the Poynting flux and $\mathbf{F}_c$ the conduction flux. Integrating this equation between the surface $S_\star$ and a spherical surface $S_\infty$ at great distance from the star $R_\infty$, we have:

\begin{equation}
\int_{S_\infty} \mathbf{F}_E \cdot d\mathbf{S} + \int_{R_\star}^{R_\infty} Q_r dV = \int_{S_\star} \mathbf{F}_E \cdot d\mathbf{S}.
\end{equation}

At the photosphere, the wind speed and sound speed are very small compared to the gravity potential and the dominant term of $\mathbf{F}_E$ are the Poynting flux $\mathbf{F}_p$ and the gravity term. Far away from the star, only the kinetic energy term remains so that we are left with:

\begin{equation}
\int_{S_\infty} \left(\frac{1}{2} v^2 \right) \rho \mathbf{v} \cdot d\mathbf{S} + \int_{R_\star}^{R_\infty} Q_r dV = \int_{S_\star} (\mathbf{F}_p - \rho \mathbf{v} \frac{GM_\star}{r})\cdot d\mathbf{S}
\label{eq:energetics1}
\end{equation}

Using mass conservation and  $GM_\star / R_\star = v_{esc}^2/2$,  we can write this equation in terms of $\dot{M}$:

\begin{equation}
\dot{M} \left(v_{esc}^2 + v_\infty^2 \right)/2 + \int_{R_\star}^{R_\infty} Q_r dV = \int_{S_\star} \mathbf{F}_p \cdot d\mathbf{S},
\label{eq:energetics}
\end{equation}

that we finally rewrite in terms of luminosity:

\begin{equation}
    L_{K} \left(1+ \frac{v_{esc}^2}{v_{\infty}^2} \right) + L_{rad} = L_p,
    \label{eq:lumin}
\end{equation}
with $L_K =\dot{M} v_\infty^2/2$, while the other terms defined by the integrals of equation \ref{eq:energetics}.

\begin{figure}
    \centering
    \includegraphics[width=3.5in]{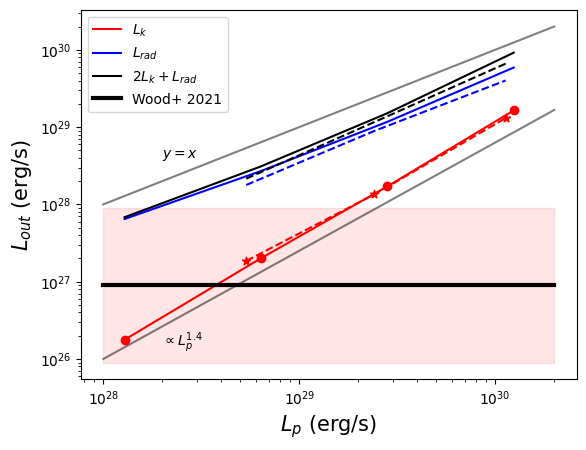}
    \caption{Illustration of the energy balance in our simulations according to equation \ref{eq:energetics}-\ref{eq:lumin}. Red lines correspond to the kinetic energy luminosity while dark blue lines to the radiated energy. The sum is shown in black and correspond to the expected input energy in light blue. Dashed lines are used when the turbulence correlation length scale $\lambda_{\star}=0.01$. The target mass loss rate (expressed in kinetic luminosity) is given by the horizontal dark line, with the reported uncertainties of figure \ref{fig:wood_law} in red shades.}
    \label{fig:energetics}
\end{figure}

Figure \ref{fig:energetics} illustrates this latter equation. Logically, the blue radiative and red kinetic terms are growing functions of the input Poynting flux. The kinetic term is a power law of the input Poynting flux with a slope of roughly 1.4. Note that this dependence differs than the exponential dependence found in \citet{Hazra2021}. The reason is that we now include the transition region with a fixed low chromospheric temperature that has a negligible influence on the energy budget and the mass loss rate \citep[see also][]{Velli2010}. The radiative losses largely dominate the energetic balance and are thus growing almost linearly with the input Poynting luminosity. This is due in large part to the closed regions of the simulations where, in steady state, radiative losses match the input energy at the coronal loop footpoints. But a significant part of the input energy is also lost in radiation in open regions. The sum of radiative and twice the kinetic terms are close, as expected, to the total input energy illustrated with the $y=x$ line (assuming that $v_\infty \sim v_{esc}$, equation \ref{eq:lumin} becomes $2 L_K + L_{rad} = L_p$). Differences are likely due to numerical integration of these terms on the first domain cells during post-processing, as the radiative losses can be very significant in these regions.

The horizontal black line shows the expected kinetic luminosity using the \citet{Wood2021} law. We used the target mass loss rate of $\dot{M} = 2.25 \times 10^{11}$ g/s, and a terminal velocity $v_\infty = 900$ km/s, consistent with the values obtained in our simulations. As we can see, the simulations with the smallest $\delta v_\star = 3, 6$ km/s are within the range of uncertainties for the Wood law shown in red shades (same as Figure \ref{fig:wood_law}). Moreover, simulations with $\delta v_\star = 12$ km/s are right at the upper bound of this region. We can thus reasonably assume that this ensemble of simulations is representative of the TRAPPIST-1 system.

Figure \ref{fig:energetics} finally shows that the global energetics varies little with the turbulence correlation scale, in agreement with equation \ref{eq:energetics}. However, as discussed at length in the previous section and as shown in Figure \ref{fig:profiles} and Figure \ref{fig:viz}, the terminal wind speed can vary significantly, more than two times the escape velocity of 529 km/s, for the highest value of the turbulence correlation length scale. 

\section{Alfvén surface and potential star-planet interactions}
\label{sec:spi}

\subsection{The TRAPPIST-1 system}

Figure \ref{fig:viz} shows in white contour the structure of the Alfvén surface in all the runs. These surfaces have the typical shape of dipolar winds, with two lobes above the poles harboring the largest values of the Alfvén radius and lower values around the equator. There the Alfvén surface joins the tip of the largest steady closed loop, between $20 R_\star$ and $30 R_\star$.

\begin{figure}
    \centering
    \includegraphics[width=3.5in]{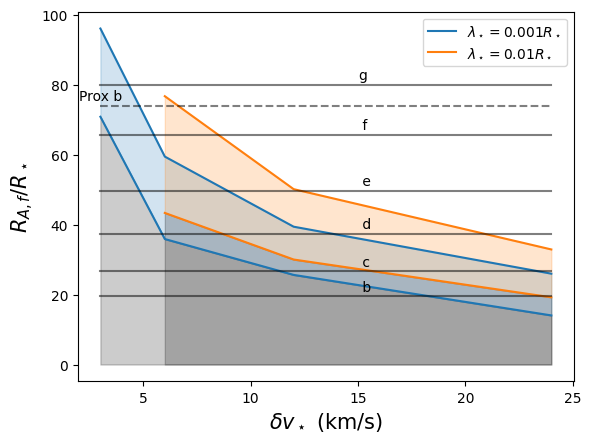}
    \caption{Extent of the fast Alfvén surface as a function of $\delta v_\star$. Grey regions mark the minimum value of the fast Alfvén surface. Colored shades show its latitudinal variability for both values of $\lambda_\star$. Horizontal black lines represent the semi-major axis of the first six planets of the TRAPPIST-1 system. For all simulations, at least one planet is within the maximal fast Alfvén radius, and can thus give birth to star-planet interaction. The dashed black line is the semi-major axis of Proxima Centauri b in stellar radii.}
    \label{fig:ras}
\end{figure}

Furthermore, for star-planet magnetic interactions to exist, having the planet lie within the fast magneto-acoustic surface of the stellar wind suffices. In practice, due to the relatively low rotation of the TRAPPIST-1 planet (the breakup ratio $f=R_\star \Omega_\star/ \sqrt{G M_\star/R_\star}=5 \times 10^{-3}$), the fast magneto-acoustic surface is very close to the Alfvén surface in the fast wind regimes. Close to the equator, the two surface are distinct, and the fast magneto-acoustic surface extends further away than the Alfvén surface. Figure \ref{fig:ras} shows the extent of the fast magneto-acoustic surface in all of our simulations (we shall call it the fast Alfvén surface/radius in the following of the manuscript). It is defined as the point where the squared fast Alfvén Mach number is unity:

\begin{equation}
    M_{A,f}^2 = \frac{2 v^2}{c_s^2+v_{A}^2+\sqrt{(c_s^2+v_{A}^2)^2-4 c_s^2 v_{A,p}^2)}}=1,
    \label{eq:ma}
\end{equation}
where $c_s$ is the sound speed, $v_A$ the Alfvén speed, and $v_{A,p}$ the poloidal Alfvén speed (without the $\varphi$ component). As already shown by this contour in Figure \ref{fig:viz}, the location of the fast Alfvén radius is variable with latitude and oscillates between a maximum value at the pole and minimum value close to the equator, at the boundary between slow and fast wind components. The interval between these two values is represented in Figure \ref{fig:ras} with colored shades, each color being associated with a value of $\lambda_\star$. For low values of $\delta v_\star$, the fast Alfvén surface can extend to very large distances, up to $90 R_\star$. 

The averaged fast Alfvén radii are reported in Table \ref{tab:res}. The fast Alfvén radius is a decreasing function of the input energy and $\delta v_\star$. This is a natural consequence of the increasing mass loss rate with increasing $\delta v_\star$. In the equation $\ref{eq:ma}$, the mass loss rate is at the numerator, and increasing either the wind speed or the mass density, will make the wind super fast-alfvénic closer to the star. Moreover, it only depends weakly on the turbulence correlation length scale. This is due to the fact that the mass loss rate itself does not depend too much on $\lambda_\star$ (see Figure \ref{fig:energetics}). Most of the effect of $\lambda_\star$ on the wind speed occurs beyond the sonic point (located at $\sim 2 R_\star$ in the fast wind regime of our simulations), and as such the acceleration due to the wave pressure is done at constant mass flux, which conserves the location of the fast Alfvén point.  

The semi-major axis of the orbit of the six first planets of the TRAPPIST-1 system are also represented in Figure \ref{fig:ras}. Looking at the range $\delta v_\star=3-12$ km/s, which is the most consistent with the expected mass loss rate obtained through the Wood law, we observe that at least planet-b must lie within the Alfvén surface. The maximum and minimum of the fast Alfvén surface can also be interpreted as a proxy for the inclination of the stellar dipole with respect to the ecliptic plane. The most favorable case correspond to the case where the dipole axis is contained in the ecliptic plane. The most unfavorable case is when the dipole axis has a slight inclination that correspond precisely to the minimum of the fast Alfvén surface, which appear unlikely.  In the most favorable case ($\delta v_\star = 3$ km/s and a strongly inclined dipole field), up to 6 planets are within the fast Alfvén zone for part of their orbit. 

Hence, we can reasonably expect the TRAPPIST-1 system to be subject to several star-planet magnetic interaction, where the orbital motion of the planet excites waves that are able to come back to the host star, creating enhanced chromospheric emission \citep{Strugarek2015, Strugarek2016, Strugarek2019}. Unfortunately, such interactions are very difficult to observe, and we do not know of any detection made for the TRAPPIST-1 system. However, our simulations are quite general and may apply to other systems. We now focus on a well known system for which such detections have been claimed. 

\subsection{SPMI detection for Proxima Cen b}

The discovery of Proxima Cen b, a possible terrestrial planet, in the habitable zone of the closest star to the Sun \citep{Anglada2016}, has triggered a renewed interest to the system. In particular, \citet{PerezTorres2021} have detected a radio signal in the $1.1-3.1$ Ghz range, compatible with an electron cyclotron maser (ECM) emission that we observe for close-in star planet or planet satellite interaction \citep[such as the Jupiter-Io system, see][]{Zarka1996}. 

For these emissions to be effectively due to an ECM process, the planet, or at least part of its orbit, must lie within the fast Alfvén surface of the star. Proxima Cen is a M5.5V type star of mass $M_\star = 0.12 M_\odot$ and radius $R_\star = 0.14 R_\odot$ \citep{Boyajian2012}. The escape velocity of Proxima Cen is $v_{esc}=565$ km/s, only slightly larger than the TRAPPIST-1 system (where $v_\mathrm{esc}=529$ km/s).  Hence, the normalization parameters are relatively close to the one used for the previous numerical study. Moreover, the magnetic flux has been estimated at $Bf = 600\pm150$ G \citep{ReinersBasri2008, Klein2021}, which is very similar to the amplitude measured for TRAPPIST-1. The rotation rate of the star is much smaller, with a period estimated at $P=83$d, and should play an even smaller role in the stellar wind dynamics than for TRAPPIST-1. It is thus reasonable to assume that the previous parameter study also applies to Proxima Cen. 

Observations of the quiescent X-ray flux of Proxima Cen fall between $4-16 \times 10^{26}$ erg/s \citep{Haisch1990, Fuhrmeister2022}. Assuming an average value $F_X = 10^{27}$ erg/s, the expected mass loss rate through the Wood law is

\begin{equation}
    \begin{aligned}
        \dot{M}_\star &= 3.8 \times 10^{11} \mbox{g.s}^{-1}\\
        &= 0.25 \dot{M}_\odot
    \end{aligned}    
\end{equation}
An upper limit for the mass loss rate has been moreover derived by \citet{Wood2001}, due to non detection. This upper limit is $\dot{M}_\star = 2.5 \times 10^{11}$ g/s or $\dot{M}_\star = 0.17 \dot{M}_\odot$, with the value used throughout this paper, $\dot{M}_\odot = 1.5 \times 10^{12}$g/s.

Given these constraints on the mass loss rate, and referring to Figure \ref{fig:energetics}, we see that the simulations with input parameter $\delta v_\star = 3-6$ km/s could well characterize the Proxima Cen wind. In Figure \ref{fig:ras}, we have plotted in dashed black the semi-major axis of the planet b, for which star-planet interaction have been detected. We see that this interval of $\delta v_\star$ is fully consistent with a planet orbiting within or very close to the fast Alfvén surface, comforting these observations. 

\begin{figure*}
    \includegraphics[width=7in]{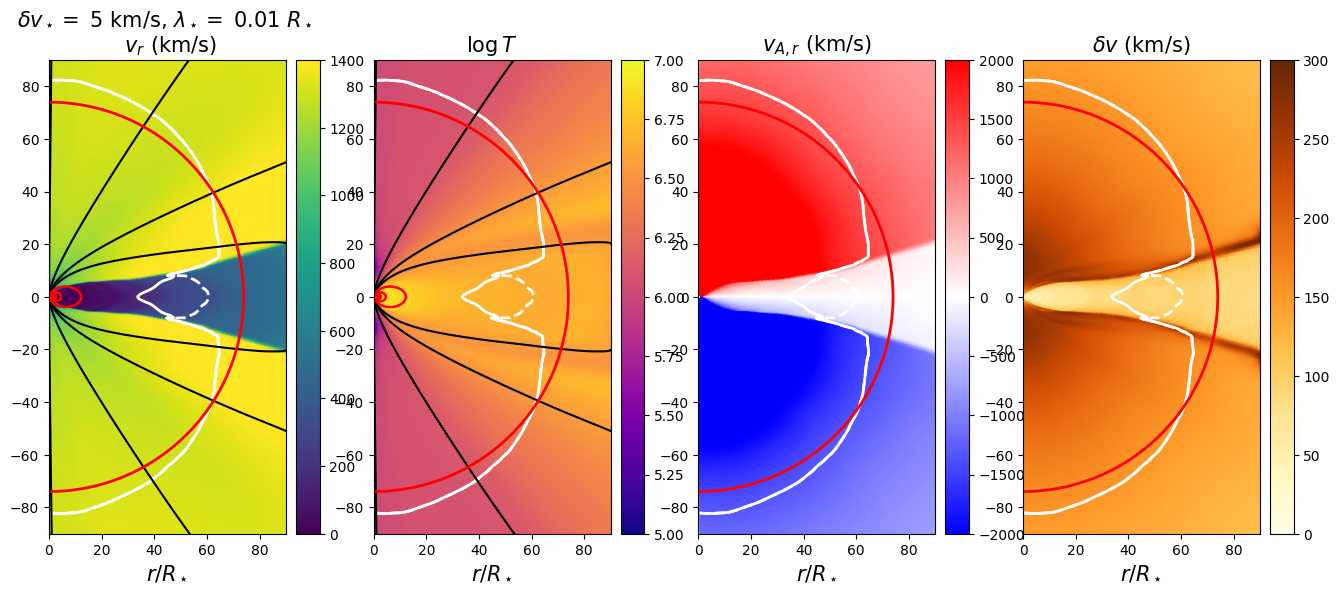}
    \caption{Same as Figure \ref{fig:viz}, for the Proxima Cen stellar parameters and $\delta v_\star=5$ km/s and $\lambda_\star=0.01 R_\star$. The mass loss rate of the simulation is at the upper bound $\dot{M}=0.17 \dot{M}_\odot$. The semi-major axis of Proxima Cen b is shown in red, within or very close to the (fast) Alfvén surface of the stellar wind.}
    \label{fig:prox}
\end{figure*}

For completeness, we perform one simulation with $\delta v_\star = 5$ km/s, $\lambda_\star = 0.01 R_\star$ and the stellar parameters of Proxima Cen. Figure \ref{fig:prox}, shows the quasi-steady solution in the fashion of Figure \ref{fig:viz}. The parameters of the simulation have been chosen so that the mass loss rate is 

\begin{equation}
    \begin{aligned}
        \dot{M}_\star &= 2.55 \times 10^{11} \mbox{g.s}^{-1},\\
        &= 0.17 \dot{M}_\odot,
    \end{aligned}    
\end{equation}
precisely at the upper limit given by \citet{Wood2001}. The orbit of the planet is below the maximum of the fast Alfvén surface, and remain close even if for a weak inclination of the stellar dipole. It thus could be at the origin of the radio signal detected by \citet{PerezTorres2021}. Note that, with a lower mass loss rate, the fast Alfvén surface would extend further, encompassing more of the orbit of Proxima Cen b.

\section{X and UV emissions}
\label{sec:xuv}

The final section of this paper is dedicated to the soft X-ray, extreme UV and far UV emissions (0.5-180 nm) modelling for M-dwarfs stellar corona. These emissions are of primary importance to assess the chemistry of planetary atmosphere and their expected lifetime due to atmospheric mass loss \citep[see, e.g.,][]{SanzForcada2011,Chadney2015}. They are, however, very difficult to observe in their entirety due to absorption processes in the interstellar medium and must rely on models constrained at a few wavelengths. 

In the following, we compare two techniques developed in recent years and compare them with observations and our simulations. We first rely on the technique introduced by \citet{Toriumi2022a, Toriumi2022b} and further developed by \citet{Namekata2023}, that derive a scaling relationship between the X and UV spectrum of the Sun and the observed unsigned magnetic flux:

\begin{equation}
    I(\nu) = I_0(\nu) + 10^{\beta(\nu)} (\Phi - \Phi_0)^{\alpha(\nu)},
\end{equation}
where the intensity $I_0$ is the spectral intensity at solar minimum, $\nu$ the wavelength, and $\Phi_0=1.18 \times 10^{23}$ Mx is the unsigned magnetic flux at solar minimum. A Table of $I_0(\nu), \alpha(\nu), \beta(\nu)$ is given in \citet{Namekata2023} (machine readable). 

\begin{figure}
    \centering
    \includegraphics[width=3.5in]{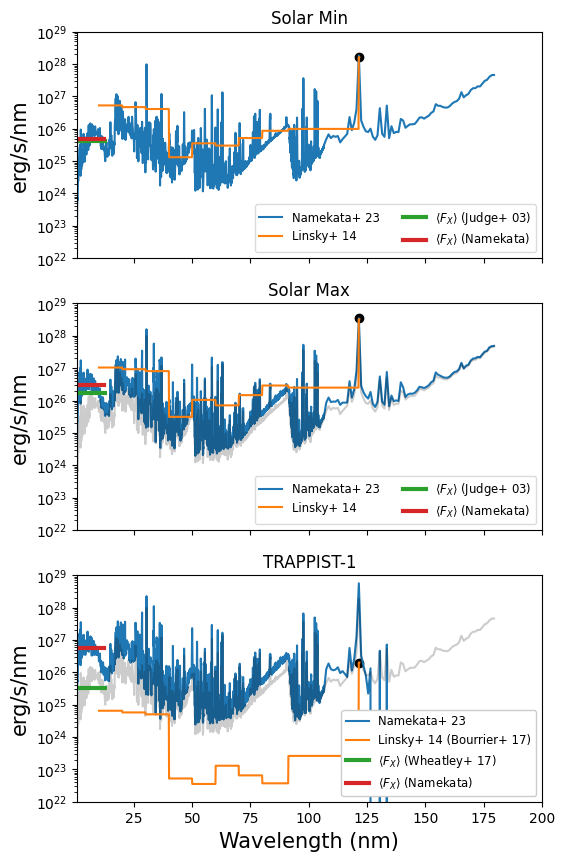}
    \caption{Reconstructed spectrum (0.5-180 nm) of the Sun (min and max) and TRAPPIST-1 using the $F-\Phi$ scaling relation of \citet{Namekata2023} in blue, and the F(EUV)-F(Ly$\alpha$) of \citet{Linsky2014} in orange. The peak of the Ly$\alpha$ line is used to constrain the \citet{Linsky2014} extrapolation. It is taken from \citet{Bourrier2017a} in the case of TRAPPIST-1. The observed X-ray flux is shown in green, while the average value from the \citet{Namekata2023} extrapolation is shown in red. The grey shaded spectrum in the middle and bottom panel is the observed solar minimum spectrum of the top panel.}
    \label{fig:xuv}
\end{figure}

For TRAPPIST-1, we can get a rough estimate of the unsigned magnetic flux, using the average value of the field, $\Phi_{\mathrm{TRAP}} = 5.3 \times 10^{23}$ Mx. Figure \ref{fig:xuv} shows in blue the spectra at solar minimum ($I_0$), solar maximum (computed with $\Phi=3.5 \times 10^{23}$ Mx) and for the TRAPPIST-1 system (note that with this technique, the estimated spectra of TRAPPIST-1 and Proxima Cen are identical). The solar black body radiation has been removed for TRAPPIST-1, which make the UV spectrum vanish soon after the Ly$\alpha$ line at 121.5 nm.

We now compare these spectra with the method developed in \citet{Linsky2014}. This work gives the average UV flux on the interval $10-117$ nm, split in nine wavelength bands, based on the flux in the Ly$\alpha$ line. Figure \ref{fig:xuv} shows this extrapolation in orange, with Ly$\alpha$ fluxes taken from \citet{Linsky2014} for the Sun, and \citet{Bourrier2017a} for TRAPPIST-1. X-ray fluxes in the range $0.1-2.4$ keV are also shown for comparison, with values from \citet{Judge2003} for the Sun, and \citet{Wheatley2017} for TRAPPIST-1. 

\begin{figure}
    \centering
    \includegraphics[width=3.5in]{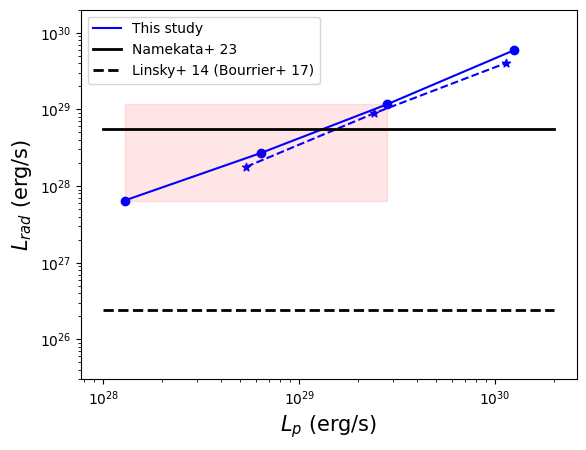}
    \caption{Radiative losses (same as Figure \ref{fig:energetics}), compared with the integral of the TRAPPIST-1 synthetic spectrum in the EUV-FUV range (10-121.7 nm) with the method of \citet{Namekata2023} and \citet{Linsky2014}. Plain and dashed blue lines correspond to $\lambda_\star = 0.001 R_\odot$ and $\lambda_\star = 0.01 R_\odot$ respectively. The red box identifies the parameter range compatible with Wood law's mass loss rate.}
    \label{fig:rad}
\end{figure}

Given the values of the unsigned magnetic flux used for the Sun and TRAPPIST-1 in the \citet{Namekata2023} estimates, the extrapolated spectra are quite close. In the case of the Sun, the formula of \citet{Linsky2014} is consistent with the magnetic flux formulation. For TRAPPIST-1, however, we obtain a much smaller UV irradiation. While the unsigned magnetic flux of the Sun and TRAPPIST-1 are close, the Ly$\alpha$ luminosity differ by two order of magnitudes. \citet{Bourrier2017a, Bourrier2017b} have measured the total Ly$\alpha$ luminosity of TRAPPIST-1 around $2 \times 10^{26}$ erg/s, while solar values range between $1.5-3.5 \times 10^{28}$ erg/s. Hence, depending on the observable used for the calibration and the extrapolation, unsigned magnetic flux or Ly$\alpha$ luminosity, we obtain two very different UV spectra. Interestingly, the X-ray flux of the Sun and TRAPPIST-1 are quite close, and while the UV spectra lie above the X-ray in the solar case, the opposite is observed for TRAPPIST-1 using \citet{Linsky2014}. This suggests that the chromosphere and transition region of very late M-type stars, from which most of the Ly$\alpha$ emission is coming, are behaving differently than solar-like. 

Direct modelling of the stellar spectra have been performed for TRAPPIST-1 \citep{Peacock2019a} and other M-type stars \citep{Peacock2019b}. These works used 1D model of the stellar atmosphere considering a fine resolution of the transition region (TR). A similar approach is not adequate with multi-D models, where the TR resolution is rarely sufficient. As explained in section \ref{sec:num}, a numerical technique is used to broaden the TR, which makes the reconstructed spectra (with CHIANTI for example) unrealistic, especially in the Ly$\alpha$, which makes up for most of the radiated energy in the FUV. Nevertheless, our treatment of the transition region guarantees that the integrated radiative losses and heating are equivalent to the fully resolved TR, and we can thus precisely constrain the energy budget of the simulations.

From the extrapolations of Figure \ref{fig:xuv}, we can compute the integrated losses in the UV range 10-121.7 nm, and compare it to the value obtained in our simulations. For both models, the Ly$\alpha$ makes up for 60 to 70\% of the integral. Note that we discarded the soft X-ray range, mostly due to small scale magnetic structures, which are not modelled in this study. While it is small compared to the Ly$\alpha$ line intensity for the Sun, it could be significant for TRAPPIST-1 (see Figure \ref{fig:xuv}). Figure \ref{fig:rad} compares the total integrated radiated emission through the term $L_{rad}$, with the integrated spectrum of both models (10-121.7 nm). The red box identifies the simulation parameter range compatible with Wood's law's mass loss rate.  With the "Sun-as-a-star" approach of \citet{Namekata2023}, the matching value for $\delta v_\star$ is around $6-12$ km/s depending on $\lambda_\star$, which is consistent with our previous analysis. The value obtained with the \citet{Linsky2014} method is much lower, due to the observed Ly$\alpha$ luminosity of \citet{Bourrier2017a}. Nevertheless, the integrated losses of our simulation, for the lowest value of the input Poynting flux, are in between these two estimates. Note that the dashed line estimate is a likely a lower bound of the integrated spectrum, as we observe the soft X-ray part of the TRAPPIST-1 spectrum to be more solar-like. Future works shall attempt to bridge the gap between the simulations and the observations in multiple lines of the X and UV spectrum.

\section{Summary}

M-dwarf stellar winds and stellar environments are subject to a lot of attention in the context of the search for habitable planets. As for most stellar winds, observations are scarce and scenarios of magnetospheric or atmospheric interaction with the host star must rely on models. In this work, we propose a strategy based on the latest measurements of mass loss rate detected through Ly$\alpha$ absorption at the astropause \citep{Wood2021}. We define an acceptable mass loss rate region between 1/10 and 10 times the fitted law of \citet{Wood2021}, that any simulation should aim for varying its input parameters. It is indeed impossible at the moment to constrain from first physical principles the input energy that is transferred from the photosphere of a given star to its corona and stellar wind. Moreover, input parameters depend on the model and the coronal heating and wind acceleration mechanisms chosen. 

We rely here on the Alfvén wave driven scenario, using the multi-D WindPredict-AW code. We perform 9 simulations varying the input parameter $\delta v_\star$ and $\lambda_\star$. We show through a global energy balance, expressed in unit of erg/s \citep[or luminosity, see][for similar approaches]{Suzuki2013, Shoda2021}, that the input energy is distributed between radiative losses and kinetic energy transferred to the wind. Throughout our parameter scan, radiative losses always strongly dominate the kinetic energy balance, and we wish to strongly underline that special care should be brought to the treatment of the transition region. In multi-D simulations, the TR has to be broadened to limit numerical cost. We use a cutoff temperature technique, inspired by \citet{Lionello2009} and refined by \citet{Johnston2020}, to adapt the TR to our numerical resolution and ensure that the radiative losses and heating are equivalent to a fully resolved TR. 

Our study then shows that the stellar wind properties do depend on the wind acceleration model and the input parameters. In particular, while the initial velocity perturbations $\delta v_\star$ controls the amplitude of the input Poynting flux, we highlight the influence of the turbulence correlation length scale on the terminal wind speed obtained in the simulations. This parameter, $\lambda_\star$, controls the height of the energy deposition by Alfvén wave turbulence in the corona and stellar wind. In the range of acceptable mass loss rates, we have been able to run simulations with $\lambda_\star= 0.001 R_\star$ and $\lambda_\star = 0.01 R_\star$. The highest value gives birth, for the same input energy, to higher wind speed than the fast solar wind, up to $1200$ km/s. The turbulence correlation length scale is usually related to the average size of granules or supergranules emerging at the top of the convective envelope, and the typical value used in solar-like simulation is $\lambda_\odot = 0.01 R_\odot$ \citep[supergranule size, see][]{Reville2020ApJS, vanderHolst2014}. Some works have claimed however that this value should be closer to the size of granules $0.001 R_\odot$ \citep{vanBallegooijen2017}. M-dwarf have deep convection zones, and can be fully convective. Differences in the small scale organization of the magnetic field of M-dwarf could lead to either one or the other parameter choice. Note also that the size of the convective structure might be uncorrelated from the size of the star. That being said, the actual function of the turbulent length scale in the simulations is to control the height of the turbulent dissipation, which needs to heat sufficiently the low corona to create enough thermal and dynamic pressure to open the very intense dipolar magnetic field and sustain a $\sim 10^6$ K corona. This may be achieved differently depending on the coronal heating model. 

In the case of TRAPPIST-1, we thus give possible values of the stellar wind speed and mass loss rates compatible with the Wood law. We find that for this range of parameters, 2 to 6 planets are orbiting within the fast Alfvén surface of the stellar wind. This suggests that strong star-planet magnetic interactions are at play in this system. For planet e and f, which lie in the usual habitable zone \citep[see][for a review of atmospheric processes]{Turbet2020}, such processes may be considered. Previous works studying TRAPPIST-1 have found different and contradictory results. It is unclear why these previous studies results are different, but our results are consistent with detection of radio emission in Proxima Cen \citep{PerezTorres2021}. 

Our parameter study can indeed be applied to this nearby system, and we found that Proxima Cen b, is also likely to orbit within the fast Alfvén surface. Our conclusions thus differ from the one of \citet{Kavanagh2021}. The latter work used the ZDI observations of \citet{Klein2021}, which obtain a large scale field strength of $\sim 200$ G, significantly smaller than the unsigned flux amplitude of \citet{ReinersBasri2008}. Whether stellar wind simulations should use one or the other can be discussed \citep[see][for a study of these effects along the solar cycle]{Hazra2021}. Using a dipole of $600$ G, we account for all the detected magnetic flux, even if part of it is concentrated at small scales and may be absent from ZDI reconstructions. We thus consider a maximal case in terms of magnetic field amplitude. The differences may be also explained by the fact that they have obtained a mass loss rate that is above the non detection limit of \citet{Wood2021}, if only slightly, and that they have considered only the Alfvén surface. Our results thus call for a reevaluation of the environment of Proxima Cen b \citep[see][]{2024arXiv240519116P}.

Finally, we attempt to use reconstructed X and UV spectra to constrain our simulations. We use two different approaches that lead to very different results. The first approach is to scale the solar X-ray and UV emissions to the total unsigned magnetic flux of a given star \citep{Namekata2023}. Using this approach, we find a spectrum for TRAPPIST-1 which is similar to the Sun during maximum of activity. The integrated losses in EUV-FUV band (10-121.7 nm) is consistent with the radiation losses obtained in our simulations, for the right mass loss rate range. However, the peak of the Ly$\alpha$ line is much higher than observed by \citet{Bourrier2017b}. The second reconstruction, based on \citet{Linsky2014} and calibrated by the observed peak in Ly$\alpha$, yields thus a much lower UV flux from TRAPPIST-1 than what is predicted by the unsigned flux approach and our simulations. None of these reconstructions have been intended for very low mass stars, the study of \citet{Linsky2014} going down to M5 stars. Moreover, our simulations are certainly closer to the "Sun-as-a-star" approach of \citet{Namekata2023}, which explains the better agreement. This discrepancy appears like an interesting friction point that should be addressed in further studies, through for example a special attention to the transition region where most of the Ly$\alpha$ emission is coming from. 

\section{Acknowledgements}

We thank the referee for insightful comments that significantly improved the manuscript. VR thanks Miguel Pérez-Torres, Vincent Bourrier, Jean-François Donati, Claire Moutou, Chen Shi and Susanna Parenti for useful discussions. JMJ, MV and NM acknowledge support from the Jet Propulsion Laboratory, California Institute of Technology, under a contract with NASA. VR, JMJ, MV, NM and LHR acknowledge support of the NASA-ROSES Exoplanet Research Program (grant number 80NM0018F0610). Simulations were performed on the IDRIS/Irene supercomputer through the GENCI grant A0150410293. AS acknowledges funding from the Programme National de Planétologie (INSU/PNP), the PLATO/CNES grant at CEA/IRFU/DAp, DIM-ACAV+ ANAIS2 project, the ERC EXOMAGNETS consolidator grant \#101125367, the DIM/Origines DynamEarth project, the MERAC fundation. ASB acknowledges support from the Programme National Soleil Terre (PNST) and the ERC WholeSun \#810218. The authors are grateful to the PLUTO development team. CHIANTI is a collaborative project involving George Mason University, the University of Michigan (USA), University of Cambridge (UK) and NASA Goddard Space Flight Center (USA).

\bibliography{sample631}
\end{document}